\documentstyle[prl,aps,epsf,multicol]{revtex}
\begin{document}
\draft

\title{Thermokinetic approach of the generalized 
Landau-Lifshitz-Gilbert equation with spin polarized current}
\author{ J.-E. Wegrowe \cite{email}}
\address{Institut de 
Physique Exp\'{e}rimentale, Ecole Polytechnique F\'{e}d\'{e}rale de 
Lausanne, CH-1015 Lausanne, Switzerland}

\maketitle

\begin{abstract}
In order to describe the recently observed effect of current induced 
magnetization reversal in magnetic nanostructures, the thermokinetic 
theory is applied to a metallic ferromagnet in contact with a 
reservoir of spin polarized conduction electrons.  The spin flip 
relaxation of the conduction electrons is described thermodynamically 
as a chemical reaction.  The diffusion equation of the chemical 
potential (or the giant magnetoresistance) and the usual 
Landau-Lifshitz-Gilbert (LLG) equation are derived from the entropy 
variation.  The expression of the conservation laws of the magnetic 
moments, including spin dependent scattering processes, leads then to 
the generalized LLG equation with spin polarized current.  The 
equation is applied to the measurements obtained on single magnetic Ni 
nanowires.
\end{abstract}

\pacs{PACS numbers: 75.40. Gd, 75.70. Pa, 72.15. Gd, 05. 70. ln, 05.60.-k \hfill}


An unexpected and spectacular effect due to spin polarization of 
conduction electrons in metallic ferromagnets, the Giant 
Magnetoresistance (GMR), appeared with the first transport studies on 
magnetic nanostructures \cite{GMR}.  The spin diffusion length of 
conduction electrons being of some few tens of nm, the relaxation of 
the conduction electron spins becomes observable when the 
magnetization can be controlled over this typical length.  Some 
predictions about the inverse effect, namely the influence of spin 
polarized current on the dynamics of the magnetization, were also 
proposed.  Berger predicted the existence of some surprising phenomena 
due to the action of spin polarized conduction electrons on domain 
walls \cite{Berger} or spin waves \cite{Berger2} in magnetic thin 
films.  Slonczewski predicted the rotation of the magnetization due to 
polarized current in multilayered systems \cite{Sloncz}, and Bazaliy 
et al.  derived from microscopic considerations a generalized 
Landau-Lifshitz-Gilbert equation \cite{Bazaliy}.  All the above 
mentioned approaches are microscopic and based on the ballistic 
approximation. 

From an experimental point of view, Freitas and Berger, Hung and 
Berger \cite{Freitas}, and Salhi and Berger \cite{Salhi} show the 
action of a high current density on domain walls in thin films.  
Recent experiments on nanostructured samples bring new evidence for 
the interpretation in terms of the action of the spin of the 
conduction electrons.  Tsoi et al.  \cite{Tsoi} show an effect of a 
high current density on spin waves generation in Co/Cu multilayers, 
Sun reported on current-driven magnetic switching in manganite 
\cite{Sun}, and Myers et al.  reported an effect of current induced 
switching in magnetic multilayer device \cite{science}.  In a recent 
work, we have evidenced an effect of current-induced magnetization 
reversal in magnetic nanowires, \cite{EPL}, \cite{soumis} where the 
reversal of the magnetization is induced by a high current at an 
applied field 20\% smaller than the normal reversal field.  The 
ballistic approximation is however difficult to justify in all these 
experiments.

A phenomenological approach based on the thermokinetic theory 
\cite{Stueckelberg}, 
\cite{Gruber} of a metallic ferromagnet in contact with a reservoir of spin 
polarized conduction electrons is proposed.  In contrast to the 
pioneering works of Johnson and Silsbee \cite{Johnson} on spin 
polarized current, the spin-flip scattering is introduced here as a 
chemical reaction.  This formulation allows the conservation laws of 
the spin polarized conduction 
electrons and the ferromagnetic order parameter to be treated 
(section I). This phenomenological approach models the different 
effects able to take place in the system by a set of coupled transport 
equations (section II). The differential equation of the chemical 
potential is derived leading to the known formula of the GMR 
(section III). For a closed ferromagnetic system without electric 
current, the common Landau-Lifshitz-Gilbert equation is derived 
(section IV).  The open system with both ferromagnetization and spin 
polarized conduction 
electrons leads to the description of the generalized 
Landau-Lifshitz-Gilbert equation with polarized current (section V).  
The consequences in terms of current-induced magnetization reversal 
are deduced.  The model is applied on the data obtained on Ni nanowires.

\section{Conservation laws}

 Spin-dependent transport processes in layered structures are 
 described on the basis 
 of the following simple picture.
  After entering in the k-th layer $ \Sigma^k $ , the incident 
 current (which was spin polarized along the axis described by the 
 unite vector $ \pm \vec{u}_{k-1}$ in the 
 layer $ \Sigma^{k-1} $) first aligns \cite{justif} along the 
 axis $\pm \vec{u}_{k}$.
  Inside the ferromagnetic layer, the population of spin up ($N_{+}$) and spin down ($N_{-}$)
is then not conserved due to spin-flip scattering, 
and some of the down spins relax to the up direction. 

In this picture, the states of the sub-system $ \Sigma^k $
 are described by the variables

\begin{equation}
(S^k, \vec{M_{0}^k}, N^k_{+},N^k_{-}, \dot{\vec{M_{0}}}^k)
\label{var}
\end{equation}

where $S^k$ is the entropy, $ \vec{M_{0}^k}\,=\, M_{s}^k \, \vec{u}_{k} $ is 
the magnetization of the layer k without current, $N^k_{\pm}$ is the number of conduction 
electrons with spin parallel to the unit vector $\pm \vec{u}_{k}$ and 
$\dot{\vec{M_{0}}}^k$ is the time derivative of the magnetization in the 
layer k without current.

If $ \vec{M_{0}^k} $ and $N^k_{\pm}$ are independent variables, the conservation 
of the magnetic momentum reads

\begin{equation}
\frac{d\vec{M^k}}{dt}\,=\, 
\dot{\vec{M_{0}}}^k\,+g\,\mu_{B}(\dot{N}^k_{+}\,-\, 
\dot{N}^k_{-})\vec{u}_{k-1}
\label{contM0}
\end{equation}	

where $\mu_{B}$ is the Bohr magneton and $g$ is the Land\'{e} factor. 

  In order to write the 
 conservation laws, the spin-flip scattering mechanism is 
 described as a chemical reaction transforming a 
 spin down into a spin up along the axis $\pm \vec{u}_{k}$. In this 
 context, the 
 reaction rate $ \dot{\Psi}^k $ is introduced as the number of chemical 
events per unit of time \cite{McGrow}, \cite{Prigo}. 
Let $I^{k \to k+1}_{+}$ and $I^{k \to k+1}_{-}$ be the current of particles
 flowing from the 
layer $ \Sigma^{k} $ to the layer $ \Sigma^{k+1}$ due 
respectively to the 
electrons with spin in the direction $ \vec{u}_{k}$ and to the 
electrons with spin in the direction $ -\vec{u}_{k}$.
The conservation of the particles is then described by
  
\begin{equation}
\left\{\begin{array}{lll}
\frac{dN^k_{+}}{dt}\,&=&\,\alpha(k-1;k)\,I^{k-1 \to k}_{+}
+\,\left(1-\alpha(k-1;k)\right)\,I^{k-1 \to k}_{-}\,-
\,I^{k \to k+1}_{+}\,-\,\dot{\Psi}^k\\
\frac{dN^k_{-}}{dt}\,&=&\,\left(1-\alpha(k-1;k)\right)\,I^{k-1 \to 
k}_{+}
+\,\alpha(k-1;k)\,I^{k-1 \to k}_{-}\,-\,I^{k \to k+1}_{-}+\dot{\Psi}^k
\label{contN1}
\end{array}\right.
\end{equation}

where $ \alpha $ is the 
spin flip probability of the alignment process \cite{DWS}. In the case of ballistic 
alignment $\alpha(k-1;k)\,=\,cos^2\left(\frac{1}{2}\theta(k-1;k)\right) $ 
with $\theta$ the angle between $\vec{u}_{k-1}$ and $\vec{u}_{k}$.
Introducing the polarized current $I_{p}$, and the normal current 
$I_{N}$ defined by

\begin{equation}
\left\{\begin{array}{lll}
I^{k-1 \to k}_{p}\,&=&\,I^{k-1 \to k}_{+}\,-\,I^{k-1 \to k}_{-}\\
I^{k-1 \to k}_{N}\,&=&\,I^{k-1 \to k}_{+}\,+\,I^{k-1 \to k}_{-}
\label{Ip}
\end{array}\right.
\end{equation}

 the Eq.~(\ref{contN1}) can be put into the form:

\begin{equation}
\left\{\begin{array}{lll}
\frac{dN^k_{+}}{dt}\,&=&\,\ I^{k-1 \to k}_{+}-
\,I^{k \to k+1}_{+}\,-\,\left[\gamma(k-1;k)I^{k-1 \to k}_{p}\, 
+\dot{\Psi}^k\right] \\
\frac{dN^k_{-}}{dt}\,&=&\,\ I^{k-1 \to k}_{-}-
\,I^{k \to k+1}_{-}\,+\,\left[\gamma(k-1;k)I^{k-1 \to k}_{p}\, +\dot{\Psi}^k\right]
\label{contN2}
\end{array}\right.
\end{equation}	

with $\gamma(k-1;k)\,=\,1\, - \,\alpha(k-1;k)\,=\,sin^2\left(\frac{1}{2}\theta(k-1;k)\right)$

Inserting Eq. (\ref{contN2}) into Eq.~(\ref{contM0}) the conservation 
of the magnetization reads: 

\begin{equation}
\frac{d\vec{M^k}}{dt}\,=\, 
\vec{\dot{M_{0}}}^k\,\,+\,
g\mu_{B}\vec{u}_{k-1}\left(I^{k-1 \to k}_{p}-\,I^{k \to k+1}_{p}\,- 
\,2\left[\gamma(k-1;k)I^{k-1 \to k}_{p}\, +\dot{\Psi}^k\right]\right)
\label{contM}
\end{equation}

The problem of the spin transfer between the polarized current and the 
magnetic layer is hence solved if the polarized current $I_{p}$ and the reaction rates
$\dot{\Psi} $ can be describe as functions of the experimentally 
accessible parameters, the current $I_{N}$, the electric field 
$E_{0}$ and the kinetic 
coefficients. This task is typically performed by the application of 
the first and second laws of the thermodynamics.

\section{Kinetic equations}

The system $\Sigma^k$ is open to heat transfer, to chemical transfer, 
and to mechanical work due to the magnetization and magnetic 
fields.
 Let us define the heat and chemical power by $P_{\phi}$ and the mechanical 
power by $P_{W}$. The first law of the thermodynamics applied to 
the layer $\Sigma^k$ gives

\begin{equation}
\frac{dE^k}{dt}\,=\,P_{\phi}^{k-1 \to k}\,-\,P_{\phi}^{k \to 
k+1}\,+\,P_{W}^{ext \to k}
\end{equation}

where $P_{W}^{ext \to k}\,=\,-\vec{H}^{ext} \cdot \dot{\vec{M}^k} $. 
Furthermore, with using the canonical definitions $T^k=\frac{\partial 
E^k}{\partial S^k}\, , \, \mu_{\pm}^k=\frac{\partial 
E^k}{\partial N_{\pm}^k}\, , \,\vec{H}^k=\frac{\partial 
E^k}{\partial \vec{M}^k}$ the energy variation is:

\begin{equation}
\frac{dE^k}{dt}\,=\,T^k \frac{dS^k}{dt}\,+\, \mu_{+}^k \frac{dN_{+}^k}{dt}\,
+\, \mu_{-}^k \frac{dN_{-}^k}{dt}\, 
- \, \vec{H}^k \frac{d\vec{M}^k}{dt}\,+\,\frac{\partial 
E^k}{\partial \dot{\vec{M}_{0}^k}}\, \frac{d \dot{\vec{M}_{0}^k}}{dt}
\label{firstPrin}
\end{equation}

In the present work we limit our analysis to the isothermal case,  
$T^k=T$. 
The entropy variation of the sub-layer is deduced from the two 
last equations, after introducing the conservation laws:

\begin{eqnarray}
T\frac{dS^k}{dt}\,& = &\, \,P_{\phi}^{k-1 \to k}\,-\,P_{\phi}^{k \to 
k+1}\,+\, (\vec{H}^k\,-\,\vec{H}^{ext}) \dot{\vec{M_{0}}^k} \nonumber\\
& &-\, \frac{1}{2} \left(A^k\,-\,2g \mu_{B}(H^{k-1}\,-H^{ext,k-1})\right)
\left(I^{k-1 \to k}_{p}-I^{k \to k+1}_{p}\,-\,2 \gamma(k-1;k) I^{k-1 
\to k}_{p}
\,-\,2 \dot{\Psi}^k\right) \nonumber\\
& & -\, \frac{1}{2} \mu_{0}(I^{k-1 \to k}_{N}-I^{k \to k+1}_{N}) 
-\,
\frac{\partial 
E^k}{\partial \dot{\vec{M}_{0}^k}} \frac{d \dot{\vec{M}_{0}^k}}{dt}
\label{entropy}
\end{eqnarray}	

where the total chemical potential is $\mu_{0}^k\,\equiv 
\,\mu_{+}^k\, +\, \mu_{-}^k$. We have furthermore
 defined $H^{k-1}\,\equiv\,\vec{H}^k \cdot \vec{u}_{k-1}$ and
$H^{ext,k}\,\equiv\, \vec{H}^{ext} \cdot \vec{u}_{k}$ . The chemical affinity of the 
reaction, defined by $ A^k\,\equiv \,
 \frac{\partial E^k}{\partial \Psi}=\mu_{+}^k\,-\,\mu_{-}^k\,$ 
has also been introduced.

The entropy being an extensive variable, the total entropy variation of
 the system is obtained by summation over the layers 1 to $\Omega$ 
 where the layer 1 is in contact to 
the left reservoir $R^l$ and the layer $\Omega$ is in contact to the 
 right reservoir $R^r$. Letting

\begin{equation}
\tilde{A}^{k}\,\equiv \, A^k \,-\, 2g \mu_{B}(H^{k-1}\,-\,H^{ext,k-1})
\label{Atilde}
\end{equation}

the total entropy variation is:

\begin{eqnarray}
T\frac{dS}{dt}\,& = &\, 
[\ldots]^{R^l \to 1} - [\ldots]^{\Omega \to R^r} \nonumber\\
& & +\, \sum_{k=1}^{\Omega}(\vec{H}^k\,-\,\vec{H}^{ext}) \, \vec{M_{0}}^k 
\,+ \sum_{k=1}^{\Omega} \left(- \frac{\partial 
E^k}{\partial \dot{\vec{M}_{0}^k}} \right) \, \frac{d \dot{\vec{M}_{0}^k}}{dt} \nonumber\\
& & +\,  \sum_{k=2}^{\Omega}\frac{1}{2} 
\left(\tilde{A}^{k-1}-\tilde{A}^k\,+\,2 \gamma(k-1;k) \, \tilde{A}^k 
\right)\,I^{k-1 \to k}_{p} \nonumber\\
& & +\, \sum_{k=2}^{\Omega} \frac{1}{2} (\mu_{0}^{k-1} \,- \, 
\mu_{0}^{k})\, I^{k-1 \to k}_{N} \,+\, \sum_{k=1}^{\Omega} 
\tilde{A}^{k} \, \dot{\Psi}^k
\label{entropytot}
\end{eqnarray}

where the two first terms in the right hand side of the equality 
stand for the heat and chemical transfer from the reservoirs to the 
system $ \Sigma $. 

The variation of entropy takes the form
 
\begin{equation}
T\frac{dS}{dt}\, = \,\sum_{i}F_{i} \dot{X}^{i} \,+\, 
P^{ext}(t)
\end{equation}

where $F_{i}$ are generalized forces and $\dot{X}^{i} $ are the conjugated generalized 
fluxes. The variation of entropy is composed by an external entropy
variation $P^{ext}(t)/T$ and by an internal entropy variation 
$dS^{int}/dt$. 

By applying the second law of thermodynamics $dS^{int}/dt \ge 0$ we
 are leading to introduce the kinetic 
coefficients $l_{\alpha \beta }$ such that 
$dS^{int}/dt=\sum_{ i}F_{i} \left( \sum_{ j} l_{ij}
F^{j} \right)$. By identification with the 
expression~(\ref{entropytot}), the kinetic equations are obtained:

\begin{eqnarray}
\left[\begin{array}{c}
I^{k-1 \to k}_{N} \\
I^{k-1 \to k}_{p} \\
\dot{\Psi}^k \\
\dot{\vec{M_{0}}^k} \\
\frac{d \dot{\vec{M}_{0}^k}}{dt} \\
\end{array}\right]
= \left[\begin{array}{ccccc}
l_{NN} & l_{Np}  & l_{Nc}  & l_{NM}  & l_{N\dot{M}} \\
l_{pN} & l_{pp}  & l_{pc}  & l_{pM}  & l_{p\dot{M}} \\
l_{cN} & l_{cp}  & l_{cc}  & l_{cM}  & l_{c\dot{M}} \\
l_{MN} & l_{Mp}  & l_{Mc}  & l_{MM}  & l_{M\dot{M}} \\
l_{\dot{M}N} & l_{\dot{M}p}   & l_{\dot{M}c}  & l_{\dot{M}M}
  & l_{\dot{M} \dot{M}} \\
\end{array}\right]
\left[\begin{array}{c}
 \frac{1}{2} (\mu_{0}^{k-1} \,- \, \mu_{0}^{k})\\
 \frac{1}{2} 
\left(\tilde{A}^{k-1}-\tilde{A}^k\,+\,2 \gamma(k-1;k) \, \tilde{A}^k 
\right) \\
 \tilde{A}^{k} \\
 \vec{H}^k\,-\,\vec{H}^{ext}\\
-\left( \frac{\partial 
E^k}{\partial \dot{\vec{M}_{0}^k}} \right)\\
\end{array}\right]
\label{kinetic}
\end{eqnarray}

The indices $N$ and $p$ stand respectively for the normal and polarized transport 
processes (see section III), the indices $c$ stands for the spin-flip scattering 
chemical reaction and the indices $M$ and $\dot{M}$ account for the dynamics of the 
magnetization (see section IV). 

The kinetic coefficients are state functions:
$l_{ij} = l_{ij }(S^k, \vec{M^k}, N^k_{+},N^k_{-}, 
\dot{\vec{M}}_{0}^k)$ and the symmetrized matrix is positive :
$ \frac{1}{2}\,\left\{l_{ji}\,+\,l_{ij}
 \right\}_{\{ij\}} \ge 0$. Furthermore, according 
to Onsager relations, the kinetic coefficients are symmetric or 
antisymmetric $l_{ij}=\pm l_{ji}$
 
We assume in the following that cross effects between electronic
 transport
 and ferromagnetic transport are negligible, so that
$l_{ij}=0$ if $ i = \{N,p,c\}$ and $j = \{M,\dot{M}\} $.
Note that a polarized current is directly produced by a non-uniform
magnetization state, through the coefficient $\gamma(k-1,k)$ 
in Eq ~(\ref{kinetic}). 

The physical meaning of the kinetic coefficients is described in 
the two following sections.

\section{Giant magnetoresistance}

We focus in this section on the first three equations of
~(\ref{kinetic}) which describes the electric transport with spin 
polarization. After performing the continuum limit, we have: 

\begin{eqnarray}
\left[\begin{array}{c}
J_{N} \\
J_{p} \\
\dot{\Psi} \\
\end{array}\right]
= \left[\begin{array}{ccc}
-L_{NN} & -L_{Np}  & L_{Nc} \\
-L_{pN} & -L_{pp}  & L_{pc} \\
-L_{Nc} & -L_{pc}  & L_{cc} \\
\end{array}\right]
\left[\begin{array}{c}
 \frac{\partial \mu_{0}}{\partial z}\\
 \frac{\partial A}{\partial z} \,-\,2 \tilde{\gamma} A \\
 {A} \\
\end{array}\right]
\label{kineticElec}
\end{eqnarray}

where $\tilde{\gamma}= \lim_{dz 
\to 0}\,\frac{d \gamma}{dz}$ and the choice of symmetric coefficients 
$L_{pN} = L_{Np}$ and $L_{Nc}=L_{cN}$ is motivated by the fact that we 
neglect in this work the direct effects of the magnetic field on the 
charge carriers \cite{spinHall}.
 
In the framework 
of the two-channel approximation \cite{twochannel}, the coupling between the two conduction 
bands is neglected (i.e. there is no cross effect \cite{approx} between 
the currents $J_{+}, 
J_{-}$, and $\dot{\Psi}$):

\begin{equation}
L_{Nc} = L_{pc}=0\,;\, L_{NN}=L_{pp}=\frac{2 \sigma_{0}}{2}
\end{equation}

where $\sigma_{0}>0$ is the mean conductivity of the two spin channels.
The conductivity asymmetry $ \beta $ of the two channels is given by 

\begin{equation}
L_{Np}\,\equiv \, -\frac{2 \sigma_{0}}{e}\,\beta
\label{beta} 
\end{equation}

Equation Eq ~(\ref{kineticElec}) leads then to the set of equations:

\begin{eqnarray}
\left[\begin{array}{c}
J_{N} \\
J_{p} \\
\dot{\Psi} \\
\end{array}\right]
= \frac{\sigma_{0}}{e}\,\left[\begin{array}{ccc}
1 & -\beta  & 0 \\
-\beta & 1  & 0 \\
0 & 0  & \frac{e}{\sigma_{0}}L_{cc} \\
\end{array}\right]
\left[\begin{array}{c}
 \frac{\partial \mu_{0}}{\partial z}\\
 \frac{\partial A}{\partial z} \,-\,2 \tilde{\gamma} A\\
 {A} \\
\end{array}\right]
\label{GMRarray}
\end{eqnarray}

with $1 \ge \beta^2$ and $L_{cc }\ge 0$.

In the stationary state, $\frac{\partial J_{N}}{\partial z}=0$, and 
assuming that $\beta$, $\sigma_{0}$ and $\gamma $ are approximately independent of z
the diffusion equation of the chemical affinity is deduced (see 
appendix):

\begin{equation}
\frac{\partial^2 A}{\partial z^2}\,=\, \left ( 
\frac{1}{l_{sf}^2} \,+\, \frac{1}{l_{DW}^2} \right ) \,A 
\, + \, k \frac{\partial A}{\partial z},
\label{diff2}
\end{equation}

where the spin diffusion length $l_{sf}$ is given by

\begin{equation}
l_{sf}\,\equiv \, \sqrt{\frac{\sigma_{0} 
\left(1-\beta^2\right)}{2eL_{cc}}},
\label{lsf}
\end{equation}

the ``Domain Wall'' diffusion length $l_{DW}$ is given by

\begin{equation}
l_{DW}\,\equiv \, \sqrt{\frac{\left(1-\beta^2\right)}{4 \tilde{\gamma} ^2}},
\label{lDW}
\end{equation}

and the parameter k is given by
\begin{equation}
k\,\equiv \, \tilde{\gamma} \frac{2 \beta^2 }{\left(1-\beta^2\right)}
\label{k}
\end{equation}

Note that the chemical affinity A is equal to the difference of the chemical 
potentials of the 
two conduction bands $A= \mu_{+}-\mu_{-}$. If we assume no rotation 
of the spin polarization axis, $\tilde{\gamma}\,= \, 0 $ (which implies antiparallel magnetic 
configuration at the interface), 
then equation Eq ~(\ref{diff2}) 
is the well known diffusion equation describing the so-called ``spin accumulation'' or ``spin depletion'' 
effect responsible for the giant magnetoresistance 
\cite{Johnson}, \cite{VanSon}, \cite{Zhang}, \cite{Valet}.
A straightforward calculation (see appendix) 
leads to the giant magnetoresistance of the interface: 
 
\begin{equation}
R^{GMR}= 2 \frac{\beta^2 }{\sigma_{0}\,(1-\beta^2)}\,l_{sf}
\label{Rgmr}
\end{equation}

\section{Landau-Lifshitz-Gilbert (LLG) equation}

In this section we focus on the magnetic transport equation without 
electric current: $J_{N}=J_{p}=0$. From Eq ~(\ref{entropy}) the entropy 
variation reduces to:

\begin{equation}
T\frac{dS}{dt}\, = \, \,P_{\phi}^{ext \to in}\,-\,P_{\phi}^{in \to 
ext}\,+\, (\vec{H}\,-\,\vec{H}^{ext}) \, \frac{d \vec{M_{0}}}{dt}+
\left(-\frac{\partial 
E}{\partial \dot{\vec{M}_{0}}} \right)\, \frac{d \dot{\vec{M}_{0}}}{dt}
\label{entropyM}
\end{equation}

So that the application of the second law of thermodynamics yields,  

\begin{eqnarray}
\left\{ \begin{array}{cc}
(\vec{H}\,-\,\vec{H}^{ext}) \, & =\, \tilde{l}_{MM} \, \frac{d 
\vec{M_{0}}}{dt}\, + 
\, \tilde{l}_{M \dot{M}} \, \frac{d \dot{\vec{M}_{0}}}{dt} \\
\left(-\frac{\partial 
E}{\partial \dot{\vec{M}_{0}}} \right)\, & = \, \tilde{l}_{\dot{M} M}\, \frac{d 
\vec{M_{0}}}{dt}\,+\, \tilde{l}_{\dot{M} \dot{M}}\,\frac{d \dot{\vec{M}_{0}}}{dt}
\end{array} \right . 
\label{kineticM}
\end{eqnarray}

where the kinetic coefficients $\tilde{l}_{\alpha \beta }$ 
are the coefficients of the inverse matrix $\{l_{\alpha \beta }\}^{-1}_{\{\alpha 
\beta\}}$.

Note that in adiabatically closed systems, $\left(-\frac{\partial 
E}{\partial \dot{\vec{M}_{0}}} \right)$, $H$ and $\frac{d 
\vec{M_{0}}}{dt}$ are state 
functions (i.e depend only of the state variables 
(S, $\vec{M}_{0}$, $\dot{\vec{M}}_{0}$), and not on $\vec{H}^{ext}$).
  Since the 
kinetic coefficients are also state functions, the first equation in ~(\ref{kineticM})
 shows hence that $\frac{d \dot{\vec{M}_{0}}}{dt} $ depends on $\vec{H}^{ext}$.
 We are then leading to impose 
$\tilde{l}_{\dot{M} \dot{M}}\,=\,0$ in order to satisfy the second equation 
in ~(\ref{kineticM}), which gives the magnetic kinetic energy 
\cite{Chanterell}. The coefficient 
$\tilde{l}_{\dot{M} M}$  can be identified to the magnetic mass, and
the first equation in ~(\ref{kineticM}) gives the total magnetic force $\vec{F}^{mag} $
 acting on the system:

\begin{equation}
\vec{F}^{mag}\,\equiv \, \tilde{l}_{M \dot{M}} \,\frac{d^2 \vec{M}_{0}}{dt^2} 
\,=\,(\vec{H}\,-\,\vec{H}^{ext}) \,-\, \tilde{l}_{MM} \, \frac{d 
\vec{M_{0}}}{dt}
\label{Newton}
\end{equation}

equation (\ref{Newton}) rewrites:

\begin{equation}
\vec{F}^{mag}\,=\,\frac{\partial }{\partial \vec{M}_{0}}(-E\,-\,\vec{H}^{ext}.\vec{M}_{0}) 
\,-\, \eta \, \frac{d
\vec{M_{0}}}{dt}
\label{Newton2}
\end{equation}

where we have identified the Gilbert
 friction coefficient \cite{Gilbert}, \cite{Suhl} to $\eta\,=\, \tilde{l}_{MM}$.

The theorem of the kinetic momentum gives the equation of the dynamics:

\begin{equation}
\frac{d \vec{M}_{0}}{dt}\,=\, \Gamma \, \left (\vec{M}_{0}\,\times 
\vec{F}^{mag} \right) \,=\,\, \Gamma 
\, \vec{M}_{0}\, \times \, \left \{ 
-\frac{\partial V}{\partial \vec{M}_{0}}\,-\, \eta \, \frac{d 
\vec{M_{0}}}{dt} \right \}
\label{dynamicsM}
\end{equation}
 
where $\Gamma$ is the gyromagnetic ratio and the magnetic Gibbs 
potential \cite{Brown} is defined by 
$ V= E \,+\, \vec{M}_{0}. \vec{H}^{ext}$. Equation (\ref{dynamicsM}) is the well 
known Gilbert equation \cite{Gilbert}, \cite{Brown}, and can be put 
into the following Landau-Lifshitz form. 
In the case of uniform magnetization we have $\vec{M}_{0}\,=\, 
M_{s}\vec{u}_{0}$, where $M_{s}$ is the saturation magnetization, Eq 
~(\ref{dynamicsM}) rewrites

\begin{equation}
\dot{{u}_{0}}\,= \,-\,g'\left(\vec{u}_{0} \times 
\vec{\nabla} V \right) -h' \vec{u}_{0} \times \left(\vec{u}_{0} \times 
\vec{\nabla} V\right)
\label{LL}
\end{equation}

where $\vec{\nabla} $ is here the gradient operator on the surface of a unite sphere.
The phenomenological parameters h' and g' 
are linked to the gyromagnetic ratio $ \Gamma $ 
 and the Gilbert damping coefficient $ \eta $ by the relations \cite{Coffey} 

\begin{eqnarray}
\left\{ \begin{array}{ccc}
 h'\,&=&\,\frac{\Gamma \alpha}{(1+\alpha^2)M_{s}} \nonumber \\
g'\,&=&\,\frac{\Gamma}{(1+\alpha^2)M_{s}} \nonumber \\
\alpha\,&=&\, \eta \Gamma M_{s}
\end{array} \right . 
\end{eqnarray}

\section{LLG equation with spin polarized current}

Let us assume an interface composed 
by an incident current $I_{p}^{i}$ of conductivity
asymmetry $ -\beta $, polarized in the direction $\vec{e}_{p}$ which 
enters in a ferromagnetic layer (F) polarized in the direction $\vec{u}_{0}$ with 
conductivity asymmetry $ \beta $. The transfer of magnetic moments is 
describes by the term $ \frac 
{dN_{p}^{F}}{dt}\,=\,\dot{N}_{+}^{F}\,-\,\dot{N}_{-}^{F} $. Equation 
(\ref{contN2}) rewrites

\begin{eqnarray}
   \frac {dN_{p}^{F}}{dt} \,&=&\, I_{p}^{i}-I_{p}^{F}\,-\,2 \gamma 
   I_{p}^{F}\, = \, \beta _{eff} I_{N}
\label{dnpdt2}
\end{eqnarray}

where $\beta _{eff} \,=\,2\beta(1-2 \gamma (1-\gamma))$ and the 
expressions of $ I_{p}^{F} $ and $ I_{p}^{i} $ are derived in the appendix.

The change of the magnetic moment of the layer due to the polarized current
is given by equation (\ref{contM}) and (\ref{LL}):

\begin{eqnarray}
\dot{{u}}\,\approx \,-\,g'\left(\vec{u}_{0} \times 
\vec{\nabla} V \right) -h' \vec{u}_{0} \times \left(\vec{u}_{0} \times 
\vec{\nabla} V\right)
+ \frac{g\mu_{B}}{M_{0}}\, \beta _{eff} \, I_{N}\,\vec{e}_{p} 
\label{PolGLL}
\end{eqnarray}					
		
where the first, second and third term in the right hand side are 
respectively the precession term (or transverse relaxation), the longitudinal 
relaxation term, and the spin transfer due to spin polarized conduction electrons.				
		
\begin{figure}
\epsfxsize=5cm
$$
\epsfbox{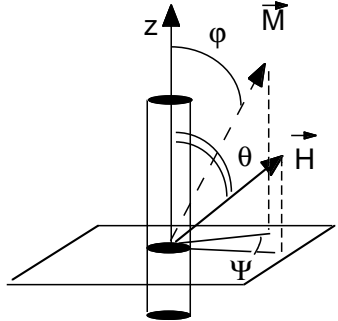}
$$
\caption{Uniform magnetization and the 
magnetic field in the case of uniaxial anisotropy.}
\label{angles}
\end{figure}

In order to estimate the effect of the injection of spin polarized 
current, the equation (\ref{PolGLL}) is applied to the case of
monodomain ferromagnet with applied field oriented at 
the angle $\theta$ from a single anisotropy axis (see Fig.~\ref{angles}).
 If the vector $\vec{u}$ 
makes an angle $\varphi$ from the 
anisotropy axis, the Gibbs energy density can be written in the following 
form \cite{Coffey}:

\begin{eqnarray}
V( \varphi \, \psi)\, = \,
K S & &\,(-cos^2 \varphi \,- \, 2h\, (cos(\theta)cos(\varphi) 
\nonumber\\
& &+\,sin(\theta)sin(\varphi)cos(\psi))
\label{potential}
\end{eqnarray}
						
 where $h=H^{ext}/H_{a}$ is the reduced applied field defined with the 
anisotropy field $H_{a}$, $K=H_{a}M_{s} $ is the anisotropy constant, 
S is the section 
and $\psi $ is the out-of-plane coordinate of the vector $\vec{u}$. 
Due to the cylindrical geometry, $\psi = 0$.
Before injecting the current, the angle $\varphi_{0}$ is given by 
the equilibrium condition $ \vec{\nabla}V=0$.

The precessional term can be neglected in (\ref{PolGLL}) (low 
frequency response and/or high damping limite \cite{Coffey}), and  $M \approx 
M_{0}$.

Experiments and samples are described in Ref. \cite{soumis},
\cite{PRL}, \cite{IEEE}.  Ni 
nanowires are obtained by the method of electrodeposition in track 
etched membrane templates.  A micro-contact is realized, and the 
magnetoresistance of a single nanowire is measured.  The wires are 
about 70 nm diameter and 6000 nm length and the magnetic energy is 
dominated by the Zeeman energy term and the shape anisotropy (or magnetostatic term),
 very close to that of an infinite cylinder.  The 
anisotropy field is calculated to be $ \mu_{0} H_{a} \approx 0.3$ T. 

The effect of the spin-polarized current was evidenced experimentally 
by injecting a 
strong current of about $2.10^7 A/cm^2$ at a fixed value
 of the external field $h = h_{sw} - \Delta h$ smaller
than the field $h_{sw}$ where the switching occurs without current.  
The magnetization switch occurs at the  angle 
$\varphi^c(\theta)$.  The 
maximum distance $\Delta h$ where the jump of the magnetization can still be 
observed corresponds then to the variation of the angle $\Delta \varphi = 
\varphi^c-\varphi_{0} $ needed to shift the magnetization up to the unstable state.

For steady states, inserting $h\,= h_{sw}\,- 
\Delta h$, $\varphi \, =\, \varphi_{c}$,
Eq.~(\ref{PolGLL}) leads to 

\begin{equation}
\Delta h\,=\, h_{sw}(\theta )\,-\, \frac{2cI_{N}\,(\vec{e}_{p}.\vec{v})
\,-\,sin(2 \varphi^c )}{sin(\varphi^c \,-\, \theta)}
\label{result}
\end{equation}										

where $\vec{v}$ 
is the polar vector perpendicular to $\vec{u}$. The parameter c is 
defined by the relation 

\begin{equation}
c\,=\,\frac{\beta _{eff}\, \hbar}{eKv_{a} \alpha}
\label{cparameter}
\end{equation}

where the activation volume $v_{a}$ of magnetization $M_{s}$ was estimated to be
$v_{a}\,\approx \,10^{-22} \,m^3$, $K\,\approx\,10^5 J/ m^3$  \cite{PRL}, and
$\, \, \beta _{eff} \, \approx \, \beta \, \approx \, 0.3 $  \cite{Piraux}, 
 $\alpha \,\approx\,0.15$  \cite{CoffeyWern}. We obtain $c\,\approx\,200 
 A^{-1}$.

All parameters in Eq.~(\ref{result}) are known if the magnetization 
reversal mode, which describes the irreversible jump, is known.  In 
some few theoretical models of magnetization reversal \cite{Aharoni}, 
the functions $H_{sw}(\theta)$ and $\varphi_{c}(\theta)$ are 
analytical.  In the framework of the present empirical approach, the 
experimental data are fitted by the relation deduced from a curling 
reversal mode in an infinite cylinder \cite{M(H)}, \cite{PRL} :

\begin{equation}
h_{sw}(\theta)\, =\,\frac{a(a+1)}{\sqrt{a^{2}+(2a+1)cos^{2}(\theta)}}
\label{curling}
\end{equation}
					
\begin{figure}
\epsfxsize=7cm
\epsfbox{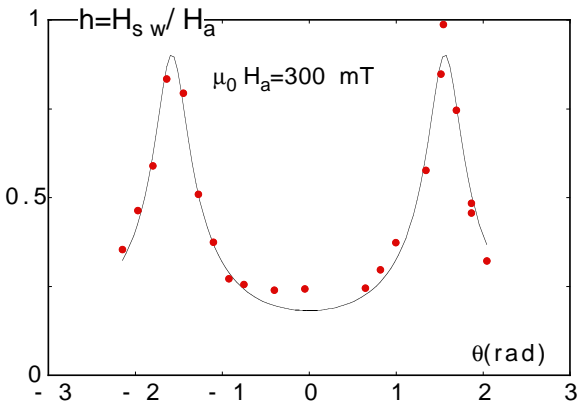}
\vspace{+0.5cm} \caption{circle: measured position of the switching 
field $H_{sw}$ for different angle of the applied field. Line: One 
parameter fit with the curling formula Eq.~(\ref{curling})}
\label{Hsw}
\end{figure}

The single adjustable parameter $a \,=\, -k \,(R_{0}/r)^2$  is defined by the 
geometrical parameter k \cite{Aharoni},
 by the exchange length $R_{0}\, =\, 20 \,nm$ ,\cite{EPL} 
and by the radius of the wire r. 
The experimental points $H_{sw}(\theta)$ are fitted in Fig.~\ref{Hsw}.

 We obtained $a\, =\, -0.15$ (which corresponds to r of about 60 nm).
 The relation between the angle of 
the applied field $\theta$ and the angle of the magnetization 
$\varphi^{c}$ is:

\begin{equation}
tan(\theta)\,=\,\frac{a+1}{a}tan(\varphi^c)
\label{curling2}
\end{equation}

\begin{figure}
\epsfxsize=7cm
$$
\epsfbox{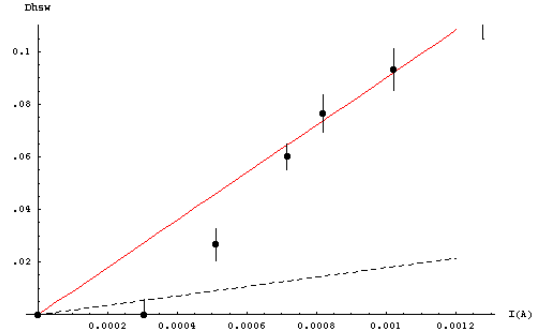}
$$
\caption{Parameter $ \Delta h\,=\, 
\Delta H_{max}(I_{e})/H_{a}$ as a function of the 
pulsed current amplitude. $\mu_{0} H_{a}\,=\,300mT$ is the anisotropy 
field. The linear fit (continuous line) gives 
$c\,\approx 190$ (see Eq.~(\ref{result})). The dashed line is 
the maximum magnetic field induced by the pulsed current}
\label{DHswIcurrent}
\end{figure}

\begin{figure}
\epsfxsize=7cm
$$
\epsfbox{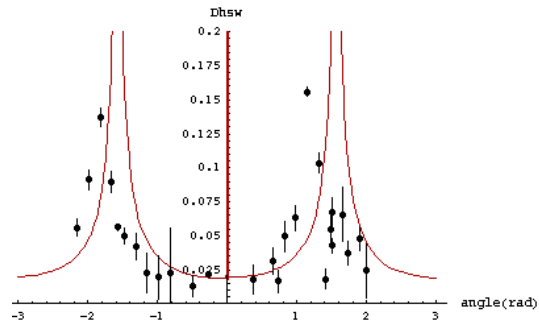}
$$
\caption{Angular dependence of the parameter $\Delta h\,=\, 
\Delta H_{max}(\theta)/H_{a}$. The curve is given by the Eq.~(\ref{result}) of the text, with 
$ c\,=\,190\, A^{-1} $}
\label{DHswI}
\end{figure}

The curve $\Delta h$, evaluated from Eq.~(\ref{result}) by numerical resolution with a 
polarization in the direction of the wire axis 
$\vec{e}_{p}.\vec{v}\,=\, sin(\varphi_{0}),$ is plotted 
in Fig.~\ref{DHswIcurrent} and Fig.~\ref{DHswI}, together with the experimental 
data. A strong discrepancy from the linear curve of $\Delta h(I_{N})$
 at small current pulses 
can be observed. Above a critical current corresponding to about 
$10^7\,A/cm^2$ the linear fit gives a parameter $c\,=\,190$, which is 
in accordance with the rough evaluation of Eq.~(\ref{cparameter}). 
This critical current below which the linear regime failed in Fig.~\ref{DHswI}
could be interpreted following Ref\cite{Sloncz} and Ref\cite{Bazaliy} as the current 
needed in order to excite spin waves or other magnetization 
inhomogeneities \cite{Kamb}. The curve given by Eq.~(\ref{result}) can then be plotted without 
adjustable parameter (Fig.~\ref{DHswI}). The divergence 
at $90^o$ is due 
to the numerical resolution of Eq.~(\ref{result}) (numerator and denominator tend to 
zero at $ \theta\,=\,90^o$ angle).

\vspace{-0.7cm}

\vspace{-0.7cm}

\section{Conclusion}

A systematic thermokinetic description of a metallic ferromagnetic 
layer open to electronic spin polarized reservoirs has been performed.  
At constant temperature, assuming the two 
current approximation and neglecting direct action of the magnetic 
field on charge carriers, five coupled transport equations account for 
the complexity of the system. The approximation of the 
explicit uncoupling of the transport processes leads to the known results about GMR 
and Landau-Lifshitz-Gilbert equations for magnetization dynamics.  
Within this approximation and on the basis of the conservation 
equation of the magnetic moment, the description of both polarized 
current and magnetization dynamics leads to a generalized 
Landau-Lifshitz-Gilbert equation.  The application of this model to 
experimental data about current-induced magnetization reversal is 
performed. 
The existence of a critical current indicates that the 
kinetics of magnetization inhomogeneity also plays an important role.
  However, the comparison with 
experimental data shows that the derived thermokinetic generalized 
Landau-Lifshitz-Gilbert equation provides a description of the basic 
mechanism responsible for the effect of 
polarized-current-induced-magnetization reversal.

\section{Acknowledgment }
This work is directly inspired by the 
synthesis performed by Professor C. Gruber on thermokinetics, and I 
thank him for his corrections and comments.  I thank Professor J.-Ph.  
Ansermet for permanent support.  Special thanks to Derek Kelly for his 
participation to the experimental work.

\section{Appendix}

The appendix is structured in three parts. In the first part (A), the 
equation of for the difference of chemical 
potentials  $A\,= \, \Delta \mu $ is derived from equation (\ref{GMRarray}) in the case of steady 
states. In part (B) the equation is applied to the simplified case of GMR or 
spin accumulation, where the polarization axis is assumed constant through the 
interface. The GMR of the interface is deduced. In part 
(C), the equation is applied in the framework of the experimental 
study of polarized-current-induced-magnetization-reversal, where 
an abrupt charge of the polarization axis occurs at the interface.

(A) 
Assuming that the kinetic coefficients which coupled the dynamics 
of the magnetization and the electric currents vanish, we obtained 
the following set of kinetic equations (\ref{GMRarray}):

\begin{eqnarray}
\left[\begin{array}{c}
J_{N} \\
J_{p} \\
\dot{\Psi} \\
\end{array}\right]
= \frac{\sigma_{0}}{e}\,\left[\begin{array}{ccc}
1 & -\beta  & 0 \\
-\beta & 1  & 0 \\
0 & 0  & \frac{e}{\sigma_{0}}L_{cc} \\
\end{array}\right]
\left[\begin{array}{c}
- \frac{\partial \mu_{0}}{\partial z}\\
 - \frac{\partial A}{\partial z}\,+\, 2 \tilde{\gamma} A \\
 {A} \\
\end{array}\right]
\label{GMRarraya}
\end{eqnarray}

In the stationary state $ \frac{\partial}{\partial z}J_{N}(z)=0 $,
 and 
assuming that $\beta$, $\sigma_{0}$ and $\gamma $ are approximately independent of z
the diffusion equation of the chemical affinity is deduced

\begin{equation}
\frac{\partial^2 \mu_{0}}{\partial z^2}\,=\, \beta \left ( \, \frac{\partial^2 
A}{\partial z^2}\,- 2 \,\frac{\partial (\tilde{\gamma} A)}{\partial z} 
\right) 
\label{diff3a}
\end{equation}

Inserting in (\ref{GMRarraya}) yields

\begin{equation}
\frac{\partial J_{p}}{\partial z}\,=\,\frac{\sigma_{0}}{e} \left( \,
(\beta^2-1)\,\frac{\partial^2 
A}{\partial z^2} \, + 2 \,\frac{\partial (\tilde{\gamma} A)}{\partial z} \right )
\label{diff4a}
\end{equation}

and by integration,

\begin{equation}
J_{p}\,=\,\frac{\sigma_{0}}{e} \left( \,
(\beta^2-1)\,\frac{\partial
A}{\partial z} \, + 2 \tilde{\gamma} \,A\, \right )
\label{jpa}
\end{equation}
where we assumed that $J_{p}(\infty)\,=\,0$ .

On the other hand, from the conservation equations we have
\begin{equation}
\frac{d}{dt}(N_{+}^k\,-\,N_{-}^k)\,=\, I_{p}^{k-1 \to k}\,-\,I_{p}^{k \to 
k+1}\,-\,2 \dot{\Psi}^k \, - \, 2 \gamma (k-1, k) \, I_{p}
\label{discret}
\end{equation}

At the continuum limit, we obtain the following relation

\begin{equation}
\frac{dn_{p}}{dt}\,=\,- \frac{\partial J_{p}}{\partial z}\,-\,2 
L_{cc}\,A \, - 2 \tilde{\gamma} \, J_{p}
\label{dnpdta}
\end{equation}

where $n_{p}$ is the density of spin polarized conduction electrons. 

Equation (\ref{dnpdta}) rewrites

\begin{equation}
\frac{\partial J_{p}}{\partial z}\,\,=\,-\,2 
L_{cc}\,A \, - 2 \tilde{\gamma} \, J_{p}\,-\,\frac{dn_{p}}{dt}
\label{djpdza}
\end{equation}

where $\frac{dn_{p}}{dt}$ is constant for steady states .  Furthermore, inside 
the ferromagnet and far away from the interface, $J_{p}$ is constant, 
whence
\begin{equation}
\frac{dn_{p}}{dt}\,=\,- \,2 
L_{cc}^{\infty} \, A_{\infty}\,=\,0
\label{dnpdtinfa}
\end{equation}

where we assumed for simplicity that 
$L_{cc}^{\infty}\,=\,0$.  Together with 
(\ref{djpdza}), (\ref{jpa}) and (\ref{diff4a}) the differential 
equation for A(z) is obtained:

\begin{equation}
\frac{\partial^2 A}{\partial z^2}\,=\, \left ( 
\frac{1}{l_{sf}^2} \,+\, \frac{1}{l_{DW}^2} \right ) \,A 
\, + \, k \frac{\partial A}{\partial z},
\label{diffAa}
\end{equation}

where the spin diffusion length $l_{sf}$ is given by

\begin{equation}
l_{sf}\,\equiv \, \sqrt{\frac{\sigma_{0} 
\left(1-\beta^2\right)}{2eL_{cc}}},
\label{lsfa}
\end{equation}

the ``Domain Wall'' diffusion length $l_{DW}$ is given by

\begin{equation}
l_{DW}\,\equiv \, \sqrt{\frac{\left(1-\beta^2\right)}{4 \tilde{\gamma} ^2}},
\label{lDWa}
\end{equation}

and the parameter k is given by
\begin{equation}
k\,\equiv \, \tilde{\gamma} \frac{2 \beta^2 }{\left(1-\beta^2\right)}
\label{k}
\end{equation}

(B) APPLICATION TO GMR.

Assuming that the polarization axis is the same for all sub-layers, we 
have $l_{DW}=0$ and the last term in (\ref{diffAa}) vanishes.

The chemical affinity obeys the diffusion equation: 
\begin{equation}
\frac{\partial^2 A}{\partial z^2}\,=\, \frac{1}{l_{sf}^2}
\,A.
\label{diff02}
\end{equation}

The chemical affinity and the total chemical potential are then

 \begin{equation}
A(z)\,=\, a\,e^{\frac{z}{l_{sf}}}\,+\,b\,e^{\frac{-z}{l_{sf}}}
\label{A(z)}
\end{equation}

\begin{equation}
\mu_{0}(z)\,=\, d\,+\,cz+\,\beta \,A(z)
\label{mu0}
\end{equation}

where $a$, $b$, $c$ and $d$ are constants.
The electric field E(z) is defined by 
$-eE(z)\equiv  \frac{\partial \mu_{0}}{\partial z}$
 so that  $c=-eE(\infty)=-eJ_{N}/\sigma_{0}$.
Under the condition of continuity of the currents of the two spin channels 
 at the interface (no surface scattering) 
: $ J_{\pm}(0^-)=J_{\pm}(0^+)$ we have $a=\frac{e l_{sf} 
\beta}{\sigma_{0}(1-\beta^2)}\,J_{N}$.
The spin polarized current on the left side of a single interface (b=0) is deduced:

\begin{equation}
J_{p}(z)\,=\,\frac{\sigma_{0}}{e}\, \left ( \frac{\partial
A}{\partial z} \,-\, \beta \, \frac{\partial
\mu_{0}}{\partial z}\right )\,=
 J_{N} \beta \,\left(e^{- \frac{|z|}{l_{sf}}}-1 \right)
\label{Jpb}
\end{equation}

The electric field $ \frac{\partial \mu_{0}}{e \partial z} \equiv 
-E(z)$ is:

\begin{equation}
E(z)\,=\,\frac{eJ_{N}}{\sigma_{0}}\, \left ( 1+\frac{\beta 
^2}{1\,-\,\beta ^2}\,e^{\frac{-|z|}{l_{sf}}} \right)
\label{Ezb}
\end{equation}

and the supplementary potential due to the spin-polarized current is

\begin{equation}
\Delta V\,=\, \int_{-\infty}^{+\infty} \left( E(z)- 
\frac{eJ_{N}}{\sigma_{0}} \right )dz\,=\, 2 \frac{\beta 
^2}{1\,-\,\beta ^2}\,l_{sf} \frac{eJ_{N}}{\sigma_{0}}
\label{dVb}
\end{equation}

from which the GMR resistance (\ref{Rgmr}) is deduced.

(C) APPLICATION TO SPIN TRANSFER

In the case of an interface composed by an incident current of 
conductivity asymmetry $ \beta $, polarized in the direction 
$\vec{e}_{p}$, entering in a ferromagnetic layer polarized in the 
direction $\vec{u}$ with conductivity asymmetry $- \beta $, the 
Equation (\ref{A(z)}) and (\ref{mu0}) still hold in the left and right 
hand sides of the interface.  However, the change of the 
polarization axis at the interface leads to modify the continuity 
equation of the current of the two spin channels $ 
J_{\pm}(0^-)=J_{\pm}(0^+) \mp \gamma J_{p}$.  The integration 
constant now reads

\begin{equation}
 a=\frac{e l_{sf} 
\beta (1-\gamma)}{\sigma_{0}(1-\beta^2)}\,J_{N} 
\label{a}
\end{equation}

and the expression of the polarized current is 

\begin{equation}
J_{p}(z)\,=
 J_{N} \beta (1 \, - \, \gamma) \,\left(e^{- \frac{|z|}{l_{sf}}}-1 \right)
\label{Jpc}
\end{equation}



\begin{thebibliography}{10}
\bibitem{justif} A proper description of this alignment process (ballistic 
vs. adiabatic alignment, or the so-called "domain-wall-scattering" 
problem) is described e.g. in the references \cite{DWS}
\end{thebibliography}
\end{document}